# Endoscopic optical coherence tomography with a flexible fiber bundle

**Lara M. Wurster,**[a*] **Laurin Ginner,**[a,b] **Abhishek Kumar,**[a] **Matthias Salas,**[a,b] **Andreas Wartak,**[a] **Rainer A. Leitgeb**[a,b]

[a]Center for Medical Physics and Biomedical Engineering, Medical University of Vienna, Währinger Gürtel 18-20/4L, 1090 Vienna, Austria
[b]Christian Doppler Laboratory for Innovative Optical Imaging and its Translation to Medicine, Medical University of Vienna, Austria

**Abstract:** We demonstrate in vivo endoscopic optical coherence tomography (OCT) imaging in the forward direction using a flexible fiber bundle. In comparison to current conventional forward looking probe schemes, our approach simplifies the endoscope design by avoiding the integration of any beam steering components in the distal probe end due to 2D scanning of a focused light beam over the proximal fiber bundle surface. We describe the challenges that arise when OCT imaging with a fiber bundle is performed, such as multimoding or cross-coupling. The performance of different fiber bundles with varying parameters such as numerical aperture, core size and core structure was consequently compared and artifacts that degrade the image quality were described in detail. Based on our findings, we propose an optimal fiber bundle design for endoscopic OCT imaging.

*Lara M. Wurster, E-mail: lara.wurster@meduniwien.ac.at

## 1  Introduction

Optical coherence tomography (OCT) is a non-invasive imaging technique that performs cross-sectional imaging of transparent, translucent or turbid media at a micrometer scale resolution.[1] Being very well established in ophthalmology,[2] OCT also holds the potential to become an important tool for in vivo imaging of internal organs by combination with endoscopy. To investigate less accessible tissues and tissues shielded by opaque media, an integration of the coherent imaging technique into endoscopic instruments proved successful.[3,4] Since cancer lesions and other types of pathologic tissue changes typically arise underneath the tissue surface, endoscopic OCT enables diseases diagnosis of internal organs at an early stage and complements the information obtained by white-light endoscopy.[5] In a best case scenario, thus, endoscopic OCT avoids the need to remove tissue samples by providing morphological tissue information at up to 2 mm in depth, therefore also being considered as 'optical biopsy'.[5] Furthermore, endoscopic OCT shows tremendous impact in imaging of regions were common biopsy cannot be performed without introducing major risks, e.g. in the vascular system or in the brain.[6]

First implementations of OCT endoscopes aimed at imaging tubular organs (e.g. blood vessels) circumferentially, by directing the light beam sideways, perpendicular to the organ wall, while performing a rotation and an axial translation of the miniaturized optics.[4]

Evidently, side imaging probes show severe limitations for imaging of hollow organs where the tissue of interest is located directly in front of the endoscope, such as the human bladder. Forward imaging endoscopy probes have been developed to address this problem, however, in general a more sophisticated instrument, housing a two-dimensional (2D) scanning unit, is required. Here, the most commonly used mechanisms integrate either a micro electro-mechanical system (MEMS) scanner or a piezo tube.[7-9] The former, however, requires a folding of the beam path, consequently increasing the overall size of the instrument. The piezo tube solution, on the other hand requires a sophisticated image reconstruction algorithm due to a spiral scan pattern which in addition is sampled non-uniformly. Previous reports have shown that the integration of a 2D scanning mechanism in a fairly small endoscope seems to be the major challenge for the development of a forward imaging probe. Furthermore, optical and biocompatibility requirements need to be considered and add complexity to the design for all the aforementioned endoscopic imaging approaches.[10]

An alternative design, capable of imaging in the forward direction and avoiding scanning at the distal end makes use of a fiber bundle (FB). Here, scanning of the light beam can be performed at the proximal end – using a pair of galvanometer mirror scanners[11] or a lens system with stage control[12], hence avoiding the need of any moving parts to be integrated into the endoscope head. However, none of the cited approaches succeeded in recording in vivo tomograms, only images of ex vivo samples were reported.

It is important to note, that individual fibers of a coherent FB in general exhibit different optical properties in comparison to using a standard SM fiber. Various factors such as core size, package density and numerical aperture (NA) strongly influence the transmission behavior and need to be carefully taken into account. Associated effects such as multimoding within each core, cross-coupling (or cross-talk) of the light among multiple cores or coupling and transmission losses might severely degrade image quality of the OCT tomogram.[11-19] Multimoding and cross-talk resemble somehow contrary effects: the former becomes more and more effective for increasing NA (large core and/or large difference in refractive index between core and cladding) and may result in ghost artifacts,[12,13] whereas cross-coupling typically arises when the NA is small and the cores are closely packed.[15] Other image quality degrading effects specific to FB imaging but independent of its NA are the pixelation artifact as well as reflections at the front and back surface of the FB. The latter can be avoided by angle polishing at both ends, therefore preventing backreflected light to be coupled back into the core. A different, simpler approach takes advantage of this backreflection of light at the distal surface of the endoscope. In so-called common-path OCT the reference and sample arm share the same optical pathway.[11] However, such a configuration restricts the use of means for detection by excluding balanced detection, which is indispensable in current swept source OCT in order to maintain sensitivity close to the shot noise limit at high imaging speeds. The pixelation artifact, on the other hand, can only be removed in a post-processing step by averaging of multiple images or by the application of one or multiple filters.[18,19] Focusing of the light at the distal end of the FB can be performed with a small GRIN lens or a lens system.[11,12] Another shortcoming of previous FB endoscopes regards their limitation in terms of flexibility. The use of rigid FBs severely restricts their successful application for imaging of hollow organs in vivo.[12]

In this paper, we present a custom developed flexible FB forward imaging OCT endoscope. Previous work on OCT with a FB used a rigid FB where images of an ex vivo sample were acquired.[12] However, the image suffered from a ghost artifact due to the multimodal behavior of the FB, although a relatively long wavelength of 1310 nm was used. We, on the contrary, want to investigate flexible FBs and moreover FBs where the image is not obscured by ghost artifacts due to multimoding. We compare the performance of two commercially available flexible FBs in addition to the rigid one used by Ref. 12 and choose the better suited to demonstrate the capabilities of our developed FB OCT endoscope by presenting, to the best of our knowledge, first in vivo imaging results. Image degradation due to on the one hand cross-coupling between cores and on the other hand improper coupling into the cores is investigated in detail. Based on our findings, an optimal FB design for endoscopic OCT imaging is eventually proposed.

## 2 Methods

A schematic drawing of the setup is shown in Fig. 1. A commercially available tunable laser source (TLS) centered at a wavelength of 1040 nm with a bandwidth of 100 nm and a sweep rate of 100 kHz (Axsun Technologies, Inc.: 1060 Swept Laser Engine) was employed and its integrated $k$-clock was used in order to avoid rescaling of the acquired spectra in post-processing. The theoretical axial resolution of the system was determined to be 4.8 μm (in air) and a sensitivity of 97.8 dB was measured close to the zero-delay with a mirror as a sample and without the FB in the sample arm. The low coherent light emitted by the TLS was guided through a fiber based symmetrical Michelson interferometer[20] which consisted of three 2×2 fiber couplers (Thorlabs, Inc.: 2×2 Wideband Fiber Optic Coupler) at a splitting ratio with regards to optical power of 50/50 each. In the sample arm, the light was collimated to a beam of 3.37 mm in diameter using a fiber collimator (C; Thorlabs, Inc.: F260APC-1064; f = 15.43 mm; NA = 0.16). The collimated beam was directed by a pair of galvo mirrors (G; Thorlabs Inc.: GVS002) to a scan lens (Fig. 1, Box 1) or scanning lens system (Fig. 1, Box 2) to achieve the best possible combination of NA and spot size fitting the respective FB's NA and core size. The scan lens (O1; Thorlabs Inc.: LSM02-BB) enabled telecentric scanning of the laser beam across the proximal end of the FB (FB3, Table 1) without the need of any additional lenses. For a different FB (FB2, Table 1) a smaller spot size was necessary and a high NA objective (O2; Nikon: 40x/0.65, WD 0.65) was chosen in combination with a 4f lens configuration (L; Thorlabs Inc.: AC254-040-B) in front of the objective to again enable telecentric scanning of the focused light beam. At the distal end of the FB the light was focused by a GRIN lens (GRINTech: GT-LFRL-180-cust-50-NC) at a working distance of approximately 670 µm with a 1:1 magnification and a size similar to the diameter of the FB itself. The resulting lateral field-of-view (FOV) is therefore equal to the FB diameter with a lateral resolution of twice the core spacing of the respective FB. The proximal end of each FB was polished at an angle of eight degrees to avoid any back reflection to interfere with the sample signal. The GRIN lens at the distal end was not angle polished. Consequently, light reflected from the GRIN lens surfaces contributed to the detected interference signal but was not obstructing the sample image features since the position of the zero delay was chosen to avoid any overlap. In order to compensate for dispersion mismatch introduced by the respective FB, additional length-matched SM fibers were inserted into the reference arm. The interference pattern was finally recorded by a dual balanced detection unit (BD, Thorlabs Inc.: PDB130C) at a frequency of up to 350 MHz at 12-bit resolution via a PCIe DAQ board (AlazarTech, Inc. ATS9360).

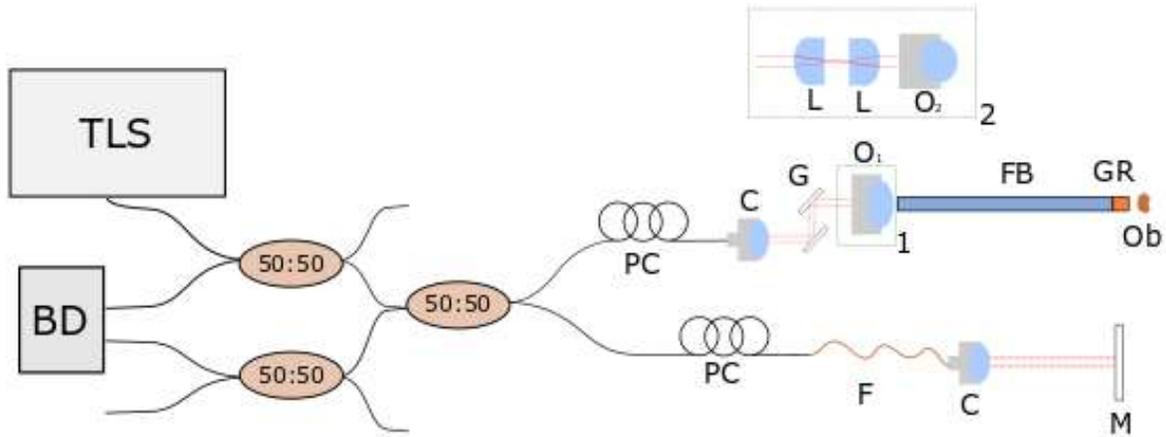

**Fig. 1** Scheme of the OCT setup to perform imaging with the custom developed FB forward imaging OCT endoscope: TLS-tunable light source: BD-balanced amplified photo-detection unit, PC-polarization control paddles, C- fiber collimator, G-galvo scanner, $O_1$-objective (Thorlabs scan lens), $O_2$-objective (Nikon), Ob-object, FB- fiber bundle, M- mirror, F-SM fiber, GR-GRIN lens, L-lens, boxes 1 and 2- different beam focusing schemes (depending on respective FB).

Three FBs differing in terms of their respective hardware specifications (cf. Table 1) were compared for applicability to perform endoscopic OCT investigations.

**Table 1** Overview of FB specifications

| Specifications | FB1 | FB2 | FB3 |
|---|---|---|---|
| Company, name | Schott Image Conduit | Fujikura FIGH-40-920G | Fujikura FIGR-10 |
| Number of fiber cores | 50419 | 40000 | 10000 |
| Diameter | 3.2 mm | 1.03 mm | 1.5 mm |
| Length | 30 cm | 30 cm | 30 cm |
| Bending radius | rigid | 100 mm | 300 mm |
| Core diameter | 9 μm | ~3 μm (irregular) | 6.1 μm |
| NA | 0.55 | 0.35 | 0.12 |
| Cut-off wavelength | 6.5 μm | 1.37 μm | 0.96 μm |

In order to estimate the properties of a fiber in terms of the number of modes that are traveling through the core the normalized wavenumber (*V*) can be determined by the following equation:

$$V = \frac{\pi \cdot d \cdot NA}{\lambda}, \tag{1}$$

where *d* denotes the diameter of the core of the individual fiber. A *V* number larger than 2.405 by definition no longer supports SM behavior.[21]

Using Eq. 1, the cut-off wavelength, which determines the maximum wavelength still supporting SM behavior, for each of the respective FBs was calculated and is stated in Table 1. Since the used light source is centered at a wavelength of 1040 nm, single mode behavior was only expected for FB3 (due to its low NA).

As mentioned previously, the lens configuration to couple light into the individual cores of the FB was chosen to match the core size and NA of the FB. Consequently, for experiments conducted with FB2 the scheme described in Fig. 1 Box 2 was used providing a spot size and NA of ~1.7 µm and 0.37, respectively. For FB3 a larger spot size but smaller NA was better suited to couple light efficiently into each core. Thus, the scheme in Fig. 1 Box 1 provided a spot size and NA of 7.0 µm and 0.09, respectively.

## 3 Results

*3.1 Detailed investigation of OCT related FB artifacts*

In general, OCT performed through FBs suffer from three artifacts: multimoding, group delay differences between fiber cores, and cross-coupling between cores. The aforementioned pixilation artifact is common to all FBS and not specific for OCT. In the following, we describe those effects individually for the three FBs given in Table 1.

*3.1.1 Multimoding*

Multimoding within fibers results in axially displaced ghost images of the structures that might severely overlap with the original structure terms. As expected, according to the analysis following Eq. 1, OCT imaging performed with FB1 suffers significantly from multimoding artifacts. Only above a wavelength of 6.5 µm SM behavior is guaranteed (Table 1). In addition, a relatively large separation between individual fibers (6 µm) results in a prominent pixelation artifact in cross-sectional as well as en face images. Still, successful OCT imaging has previously been demonstrated by Xie et al. by using the same but longer FB in combination with a 1310 nm wavelength light source[12]. However, even for the employed longer FB one ghost image remained within the imaging range. Given the strong multimoding at 1060nm as well as the fact, that the FB is rigid, we discarded FB1 from the subsequent investigations.

Artifacts due to multimoding are not expected in FB3, due to the very small NA, resulting, as mentioned previously, in a low cut-off wavelength of 960 nm. Consequently, only a single mode contributes to the detected signal. Also for FB2, multiple modes cannot be observed in the cross-sectional images, despite the cut-off wavelength of 1.37 µm was determined. Here, a possible explanation is that additional modes are shifted out of the imaging range due to the FB's length. This observation and assumption is confirmed by previously published results (cf. Ref. 13) where a large mode separation (4.4 mm for 0.8 µm wavelength) for a FB with a length of 30 cm was estimated.

*3.1.2 Group delay variations*

Another complication that can occur when imaging with a FB is performed is caused by group delay variations. The group delay defines the axial position of the measured structure with an OCT A-Scan. Variations can be introduced mainly by an irregular core size, as in FB2. They are also influenced by cross-coupling with light from neighboring cores, when imaging with a FB with a very low NA (as for FB3) is performed. To further investigate the influence of difference in core size and low NA on OCT tomograms, a more detailed examination was performed and is described in the following.

In an ideal scenario a cross-sectional OCT image of a flat surface results in a line with the maximum intensity pixel in each A-Scan at the same axial position. This is however not the case when imaging with a FB is performed. To visualize these effects, a mirror surface was imaged with both flexible FBs. After reconstruction of the OCT tomograms, a graph plotting the maximum intensity position of each A-scan within the respective tomograms with respect to a linear fit - corresponding to the theoretically flat mirror surface – is generated (cf. Fig.2). In Fig. 2 (a) the effect of the irregular core size of FB2 can be observed through large deviations of the individual intensity maxima from the linear fit. The graph for FB3 (Fig. 2 (b)), shows much smaller deviation from the linear fit. The remaining axial offsets of the intensity maxima can be attributed either to cross-coupling, ineffective coupling or a combination of both. Fig. 2 shows an exemplary single B-Scan of FB2 and FB3, respectively. Taking 20 consecutive B-Scans into account the following results in terms of the deviation of the linear fit are obtained: FB2: mean deviation of 196,4 µm (standard deviation: 167 µm); FB3: mean deviation of 23 µm (standard deviation 23,5 µm). These results confirm that the difference of each core in terms of shape and size in FB2 degrades the integrity of OCT images. Light travels a slightly different path-length in each core, consequently introducing axial shifts in the A-Scan. In theory, the referred shifts could be corrected in a post processing step by imaging a reference sample (i.e. flat mirror) in a calibration step. The measured shifts are then later applied to the image of the desired object. However, in practice, this calibration technique suffered from imprecision in repositioning of the galvo scanners and small mechanical fluctuations of the FB position relative to the scanners. This resulted in inconsistent coupling into the cores of the FB, ultimately leading to varying axial shifts. An alternative method is to use the reflex from the flat GRIN lens surface at the distal end that produces an OCT signal above the actual structure terms. Such calibration works fine for FB3, where the fluctuations remain small (cf. Fig. 2 (b)). For FB2 the fluctuations only for the GRIN lens reflex consume already >1 mm of the depth range according to Fig.2 (a). With the axial structure extension consuming typically 2 mm of depth range, the necessary axial range adds rapidly up to more than 5 mm exceeding our and most FDOCT ranging settings. As demonstrated in Fig. 2 (b) characteristics of FB3 introduce a much smaller shift in each A-Scan.

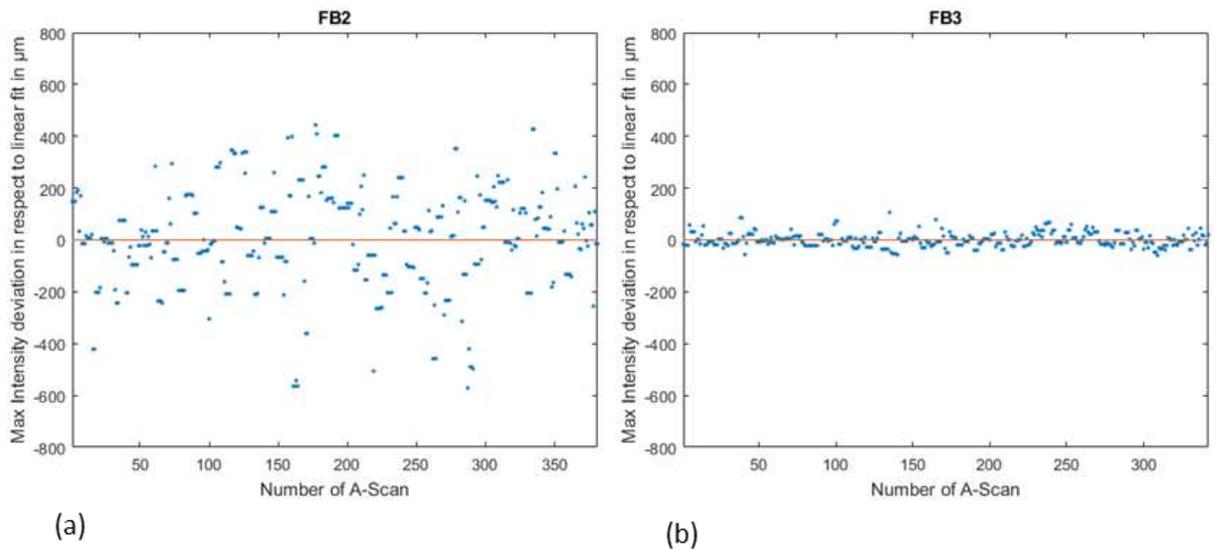

**Fig. 2** Performance of FB2 and FB3 by plotting the difference in axial location of the maximum intensity position along a cross-sectional image of a flat surface (mirror). (a) Result for a single B-Scan acquired with FB2. (b) Result for a single B-Scan acquired with FB3.

*3.1.3 Cross-coupling*

To investigate the effect of cross-coupling on OCT images the performance of FB2 and FB3 when imaging a resolution target (RT) is compared (Thorlabs Resolution Target R2L2S1P, 29 cycles/mm). To do so, en face mean intensity projections of the recorded volumetric data sets are generated. In Fig. 3(a) the individual cores of FB2 are visible and the number and bars are clearly defined. This leads to the assumption that cross-coupling is avoided when light travels through FB2, due to the higher NA and therefore higher difference in refractive index of core to cladding. It should be noted that the FOV was limited by the objective at the scanning side to ~500μm and the best resolution is only obtained in the central region of the image due to the small depth of field and angled proximal surface of the FB. The en face projection acquired using FB3 on the other hand shows a degradation of image quality (cf. Fig. 3(b)). The same number and bars of the resolution target as shown in Fig. 3 (a) are imaged with a larger FOV and the individual cores are clearly visible. However, the bars and numbers of the resolution target are hardly identifiable. Consequently, it can be assumed that cross-coupling takes place and does not only influence the axial information of a cross-sectional image (Fig. 2 (b)) but also the en face information of a 3D OCT. The information is smeared out due to the light not being confined to a single core but transmitted also to neighboring cores. Furthermore, it can be observed that light is not equally coupled into each individual core and in some regions no light seems to be transmitted through the FB at all, possibly due to contaminated fiber facets.

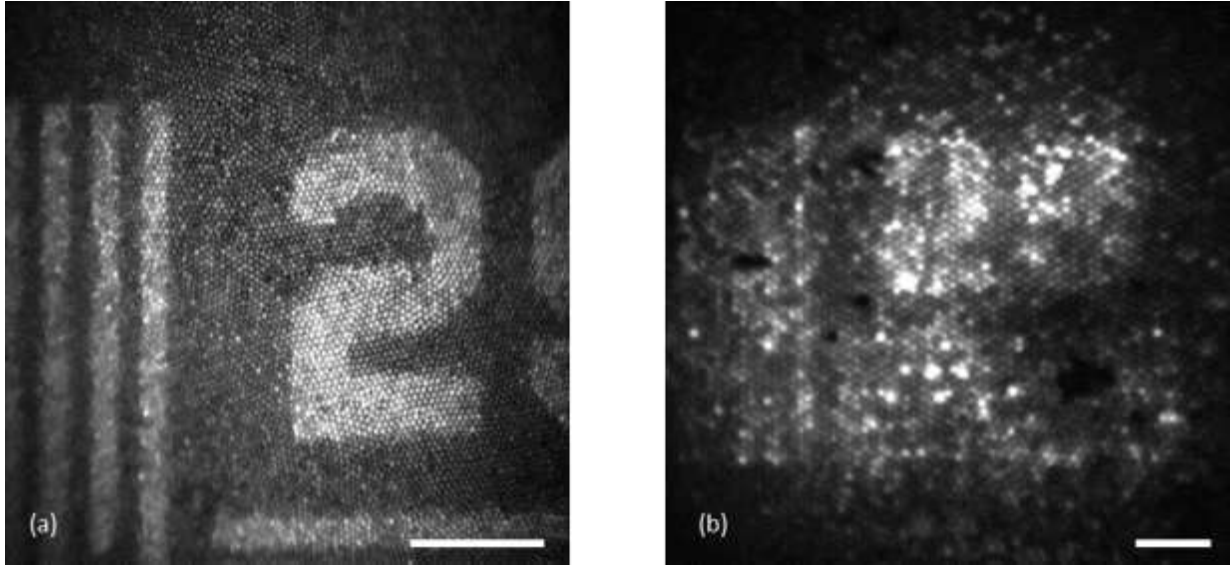

**Fig. 3** En face projection of a resolution target imaged with FB2 and FB3 to compare their performance. (a) En face projection acquired with FB2. (b) En face projection acquired with FB3. Scale bars: 100μm.

Based on the above results we conclude that imaging with FB2 can only be performed at a rather restricted performance due to the irregular core size being very critical in terms of OCT imaging. Although FB3 showed a sever degradation in image quality in the en face image we conclude that OCT can still be performed since axial shifts of the signal, caused by cross-coupling, were limited to a very small region. Consequently, further measurements and experiments were conducted with FB3.

To examine the effect of cross-coupling and a non-perfect positioning of the proximal scanning beam with regards to the center of one individual core in more detail, additional experiments were performed: Figure 4 (a) depicts a peak obtained by optimizing the position of the focused beam with respect to the center of one individual core (blue curve), while illuminating a mirror in the sample arm. Therefore, the FB was slightly moved transversally to achieve a steep and narrow peak in the OCT A-Scan. For this optimal coupling an axial resolution of ~8 μm was measured and the difference to the theoretical 4.8 μm is possibly due to spurious cross-coupling and dispersion mismatch. Thereupon, the FB was slightly moved in lateral direction resulting in the disappearance of the main peak. In this configuration multiple smaller peaks appeared in the OCT A-Scan (cf. Fig. 4 (a) red curve). After this measurement was performed the mirror in the sample arm was replaced by two lenses (Thorlabs Inc.: AC254-030-B, AC254-200-B) to image and magnify the output of the FB onto a 2D CMOS camera (Photonfocus, MV1-D2080-160-CL, 2080 x 2080 pixel resolution, 8 µm x 8 µm pixel size). Using this setup, Fig. 4 (b), (c) and (d) were obtained. Figure 4 (b) and (c) were generated using the OCT light source, comparing the following two scenarios: (i) the OCT beam is aimed directly at the center of one core (cf. Fig. 4 (b)); (ii) the OCT beam is not aimed at a single core but at a region between multiple cores. The latter is further referred to as extrinsic cross-coupling. Obviously, even light, which is perfectly coupled into a single core, is emitted by several cores at the distal end. These results confirm that intrinsic cross-coupling does take place. In a real scenario, the scanning beam is moved across the FB as indicated in Fig. 4 (d). This image of the facet was obtained, by shining white light into the proximal end of the FB. The green line indicates the scanning beam to generate an OCT B-Scan. It becomes immediately obvious, that the scanned light beam does not always coincide with the central position of a core, resulting in extrinsic cross-coupling and severe degradation of the OCT signal in such cases. Furthermore, Fig. 4 (d) shows that fibers are in general arranged linearly but the orientation of these lines can be observed to vary within the FB. Therefore, an alignment of the FB's core structure in respect to the scanning beam implicates an intrinsic challenge.

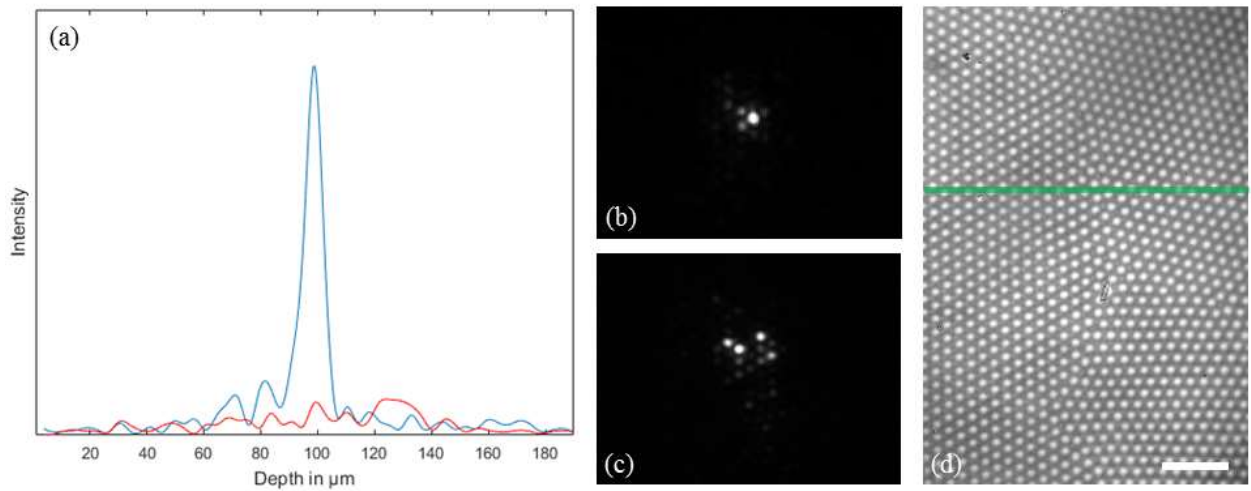

**Fig. 4** Illustration of the coupling problem when using FB3. (a) Graph obtained by placing a mirror at the sample position and carefully moving the FB to obtain a narrow peak (blue) (indicating that light hits the center of a FB's core) and multiple smaller peaks (red) when moving the FB slightly laterally. (b) Image of the output of the FB when most of the light is coupled into a single core (corresponding to the blue curve in (a)) and (c) multiple cores (corresponding to the red curve in (a)). (d) Image of the FB's facet to show the variety in arrangement of the fibers and indicating (green line) a scanning beam for an OCT B-Scan. Scale bar: 50μm.

*3.2    In vivo measurements*

Figure 5 shows images of a human finger nailbed and finger pad acquired with FB3. Informed consent from the healthy volunteer was obtained prior to the measurements and the study was in agreement with the tenets of the Declaration of Helsinki and approved by the institutional ethics committee. One acquired 3D-stack contains 700 B-Scans with 700 A-Scans each and 680 depth pixels. A sensitivity of 86 dB was measured (when carefully coupled into the center of a single core of the FB). Gel was applied onto the sample for reasons of refractive index matching, thus increasing the penetration depth by decreasing specular reflections at the air/tissue interface. In order to compensate for the axial shifts in each A-Scan a calibration with the flat GRIN lens surface was applied to all images as mentioned in Sec. 3.1.2.

Figures 5 (a) - (e) show several representations of the finger nail bed. A rendered 3D visualization is depicted in Fig. 5 (a) and single frames from positions indicated by the colored frames in (a) are shown in Fig. 5 (b) – (d). Figure 5 (e) shows an average projection of 40 consecutive scans in the proximity of the tomogram shown in Fig. 5 (b). Averaging of multiple B-scans acquired at the exact same location does not necessarily improve the contrast in a FB based system, since it will only enhance the core structure. However, by moving one axis of the galvo scanner over a small region only, while keeping the deflection of the other axis, averaging over multiple cores can be performed. An average of 250 images of a finger pad is shown in Fig. 5 (f).

Obviously, if compared to images acquired with a standard bench-top OCT system the images presented in Fig. 5 suffer from aforementioned shortcomings (cf. Sec. 3.1) that the use of a FB for endoscopic OCT imaging brings along. However, morphological structures such as the nail plate, dermis, epidermis and sweat glands of the human finger can be identified.

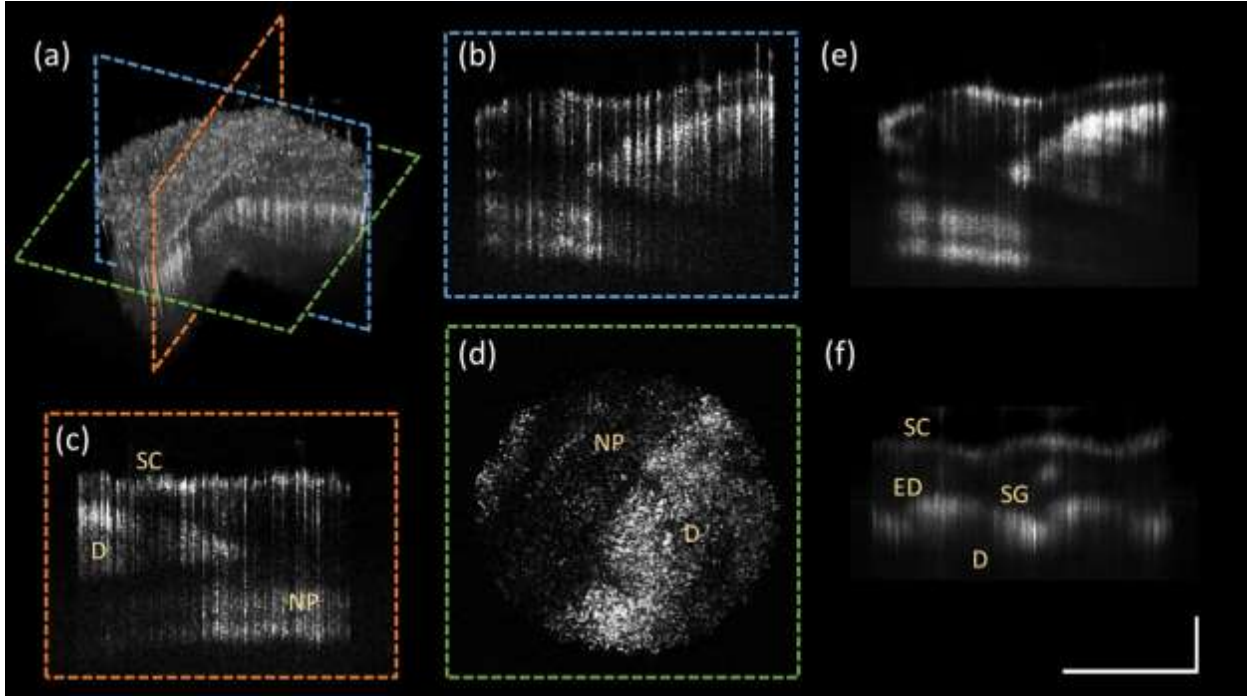

**Fig. 5** In vivo imaging results of the nailbed of a human finger (a)-(e) and finger pad (f). (a) 3D representation of the nail bed. (c)-(d) Single frames from positions indicated by the colored boxes in (a). (e) 40 times averaged image of consecutive cross-sectional images of the nailbed. (f) 250 times averaged image of a human finger where a sweat gland can be identified. SC-stratum corneum, D-dermis, NP-nail plate, ED-epidermis, SG-sweat gland. Scale bars 500 μm for all images.

## 4   Discussion

In this paper we presented first in vivo OCT imaging results obtained with a flexible FB where scanning was performed at the proximal end. This has the advantage that any beam scanning device can be avoided at the distal end and imaging in the forward direction is feasible. The presented approach implicates image degrading shortcomings related to well-known characteristics of FBs such as cross-coupling or multimoding. Experiments were conducted with three different FBs. However, FB1 was quickly excluded from further experiments after confirmation was obtained that the multiple modes traveling through this bundle at the 1040 nm wavelength range would superimpose the image information. However, we want to reference to Xie et al who imaged ex vivo tissue using the same FB but at a longer wavelength (1310 nm) with only one remaining ghost image[12]. Investigations with different phantom measurements using FB2 showed that the variation in core size is a great challenge for OCT. However, this FB has successfully been used in combination with a coherent anti-Stokes Raman scattering (CARS) setup to image rabbit aorta ex vivo.[22] FB3 has a very low NA and consequently suffers from cross-coupling that becomes mostly prominent when looking at an en face projection image e.g. from a resolution target. Less prominent but still visible is the effect when looking at a cross-sectional image where it causes the detected signal to have a slight shift for each A-Scan in the axial direction and furthermore a decline of the axial resolution is observed. Additionally, the axial resolution and shape of the signal in an A-Scan depends on the position of the scanned beam in respect to the FB core. One option to avoid that problem is to use a reference image as shown in Fig. 4 (d) where the individual cores are visible and consequently create a map to control the driving signals of the galvo scanners in a way that addresses each central point of a core individually.

Even though all the three applied FBs showed severe limitations in one or the other way, first in vivo imaging results could be obtained using FB3. Our observations allow the conclusion that the ideal FB for endoscopic OCT imaging would need particular specifications in terms of core size and NA to on the one hand avoid multimoding but on the other hand prohibit cross-coupling of the transmitted light. It can be estimated that, when taking an NA of 0.35 and a central wavelength of 1040 nm into account, a very small but regular core size of roughly ~2.3 µm would be necessary to fulfill the above requirements. When using a longer wavelength, such as 1310 nm, a slightly larger core size of ~3µm would also be acceptable. To the best of our knowledge such a FB is currently not on the market but would in theory enable endoscopic OCT imaging at an improved performance.

## 5 Conclusion

In this work, we presented a flexible forward imaging FB endoscope for OCT at 1040 nm central wavelength. We identified current limitations that arise when coherent imaging with a FB is conducted and compared the performance of three different FBs. In addition to ex vivo phantom measurements, we present 2D and 3D in vivo imaging results of a human finger. Since a theoretically ideal FB for OCT imaging is currently not available on the market, the required specifications of such a FB to achieve improved imaging performance were proposed.

*Disclosures*

The authors declare that there are no conflicts of interest related to this article.

*Acknowledgments*

The authors acknowledge financial support from the European Union project MIB (Horizon2020, contract no. 667933) and the Austrian Federal Ministry of Science, Research and Economy. Furthermore, we would like to thank Bernhard Messerschmidt from GRINTECH GmbH, Jena, Germany, for providing us with GRIN lenses and Sebastian Dochow and Iwan Schie for their help setting up a first OCT setup and their supply of FBs. Furthermore, we want to thank Marco Augustin, Bernhard Baumann, Danielle J. Harper, Pablo Eugui, Antonia Lichtenegger, Fabian Placzek, Elisabet Rank, Mikael Erkkilä and Florian Beer for helpful discussions and assistance.

*References*

1. D. Huang et al., "Optical coherence tomography," *Science* **254**, 1178-1181 (1991) [doi:10.1126/science.1957169].

2. W. Drexler and J. G. Fujimoto, "State-of-the-art retinal optical coherence tomography," *Prog. Retin. Eye Res.* **27**(1), 45-88 (2008) [doi:10.1016/j.preteyeres.2007.07.005].

3. G. J. Tearney et al., "In vivo endoscopic optical biopsy with optical coherence tomography," *Science* **276**(5321), 2037-2039 (1997) [doi:10.1126/science.276.5321.2037].

4. G. J. Tearney et al., "Scanning single-mode fiber optic catheter–endoscope for optical coherence tomography," *Opt. Lett.* **21**(7), 543-545 (1996) [doi:10.1364/OL.21.000543].

5. J. G. Fujimoto et al., "Optical biopsy and imaging using optical coherence tomography," *Nat. Med.* **1**(9), 970-972 (1995) [doi:10.1038/nm0995-970].

6. M. J. Gora et al., "Endoscopic optical coherence tomography: technologies and clinical applications [Invited]," *Biomed. Opt. Express* **8(5)**, 2405-2444 (2017) [doi:10.1364/BOE.8.002405].

7. S. A. Boppart et al., "Forward-imaging instruments for optical coherence tomography," *Opt. Lett.* **22**(21), 1618-1620 (1997) [doi:10.1364/OL.22.001618].

8. Y. Pan, H. Xie, and G. K. Fedder, "Endoscopic optical coherence tomography based on a microelectromechanical mirror," *Opt. Lett.* **26**(24), 1966-1968 (2001) [doi:10.1364/OL.26.001966].

9. L. Huo et al., "Forward-viewing resonant fiber-optic scanning endoscope of appropriate scanning speed for 3D OCT imaging," *Opt. Express* **18**(14), 14375-84, (2010) [doi:10.1364/OE.18.014375].

10. W. Drexler and J. G. Fujimoto, *Optical Coherence Tomography: Technology and Applications*, Springer, Berlin Heidelberg, (2008) [doi:10.1007/978-3-540-77550-8].

11. H. D. Ford and R. P. Tatam, "Coherent fibre bundles in full-field swept-source OCT," *Proc. SPIE* **7168** (2009).

12. T. Xie et al., "Fiber-optic-bundle-based optical coherence tomography," *Opt. Lett.* **30**(14), 1803-1805 (2005) [doi:10.1364/OL.30.001803].

13. J. H. Han and J. U. Kang, "Effect of multimodal coupling in imaging micro-endoscopic fiber bundle on optical coherence tomography," *Appl. Phys. B* **106**(3), 635-643 (2012) [doi:10.1007/s00340-011-4847-y].

14. M. D. Risi et al., "Analysis of multimode fiber bundles for endoscopic spectral-domain optical coherence tomography," *Appl. Opt*. **54**(1), 101-113 (2015) [doi:10.1364/AO.54.000101].

15. X. Chen, K. L. Reichenbach, and C. Xu, "Experimental and theoretical analysis of core-to-core coupling on fiber bundle imaging," *Opt. Express* **16**(26), 21598-21607 (2008) [doi:10.1364/OE.16.021598].

16. J. A. Udovich et al., "Spectral background and transmission characteristics of fiber optic imaging bundles," *Appl. Opt*. **47**(25), 4560-4568 (2008) [doi:10.1364/AO.47.004560].

17. H. D. Ford and R. P. Tatam, "Characterization of optical fiber imaging bundles for swept-source optical coherence tomography," *Appl. Opt*. **50**(5), 627-640 (2011) [doi:10.1364/AO.50.000627].

18. J. H. Han, J. Lee, and J. U. Kang, "Pixelation effect removal from fiber bundle probe based optical coherence tomography imaging," *Opt. Express* **18**(7), 7427-7439 (2010) [doi:10.1364/OE.18.007427].

19. C. Y. Lee and J. H. Han, "Elimination of honeycomb patterns in fiber bundle imaging by a superimposition method," *Opt. Lett*. **38**(12), 2023-2025 (2013) [doi:10.1364/OL.38.002023].

20. T. Klein, W. Wieser, C. M. Eigenwillig, B. R. Biedermann, and R. Huber, "Megahertz OCT for ultrawide-field retinal imaging with a 1050nm Fourier domain mode-locked laser," *Opt. Express* **19**(4), 3044-3062 (2011) [doi:10.1364/OE.19.003044].

21. B. E. A. Saleh and M. C. Teich, *Fundamentals of Photonics,* Chapter 9, Wiley, (2007).

22. A. Lukic, S. Dochow, O. Chernavskaia, I. Latka, C. Matthaus, A. Schwuchow, M. Schmitt, and J. Popp, "Fiber probe for nonlinear imaging applications," *J. Biophotonics* **9**(1-2), 138-143 (2016) [doi:/10.1002/jbio.201500010].

**Lara M. Wurster** has received her BSc and MSc degrees in Biomedical Engineering from the University of Luebeck, Germany. Currently, she is enrolled as a doctoral student in Medical Physics at the Center for Medical Physics and Biomedical Engineering at the Medical University of Vienna, Austria. Her research mainly focuses on the design and development of endoscope probes for OCT imaging of internal organs.

Biographies and photographs for the other authors are not available.

**Caption List**

**Fig. 1** Scheme of the OCT setup to perform imaging with the custom developed FB forward imaging OCT endoscope: TLS-tunable light source: BD-balanced amplified photo-detection unit, PC-polarization control paddles, C- fiber collimator, G-galvo scanner, $O_1$-objective (Thorlabs scan lens), $O_2$-objective (Nikon), Ob-object, FB- fiber bundle, M- mirror, F-SM fiber, GR-GRIN lens, L-lens, boxes 1 and 2- different beam focusing schemes (depending on respective FB).

**Table 1** Overview of FB specifications

**Fig. 2** Performance of FB2 and FB3 by plotting the difference in axial location of the maximum intensity position along a cross-sectional image of a flat surface (mirror). (a) Result for a single B-Scan acquired with FB2. (b) Result for a single B-Scan acquired with FB3.

**Fig. 3** En face projection of a resolution target imaged with FB2 and FB3 to compare their performance. (a) En face projection acquired with FB2. (b) En face projection acquired with FB3. Scale bars: 100μm.

**Fig. 4** Illustration of the coupling problem when using FB3. (a) Graph obtained by placing a mirror at the sample position and carefully moving the FB to obtain a narrow peak (blue)

(indicating that light hits the center of a FB's core) and multiple smaller peaks (red) when moving the FB slightly laterally. (b) Image of the output of the FB when most of the light is coupled into a single core (corresponding to the blue curve in (a)) and (c) multiple cores (corresponding to the red curve in (a)). (d) Image of the FB's facet to show the variety in arrangement of the fibers and indicating (green line) a scanning beam for an OCT B-Scan. Scale bar: 50µm.

**Fig. 5** In vivo imaging results of the nailbed of a human finger (a)-(e) and finger pad (f). (a) 3D representation of the nail bed. (c)-(d) Single frames from positions indicated by the colored boxes in (a). (e) 40 times averaged image of consecutive cross-sectional images of the nailbed. (f) 250 times averaged image of a human finger where a sweat gland can be identified. SC-stratum corneum, D-dermis, NP-nail plate, ED-epidermis, SG-sweat gland. Scale bars 500 µm for all images.